\documentclass{aa}  

\usepackage{graphicx}
\usepackage{longtable,lscape}
\usepackage[authoryear]{natbib}
\bibpunct{(}{)}{;}{a}{}{,}
\usepackage{txfonts}

\begin{document}

    \title{The eclipsing, double-lined, Of supergiant binary Cyg OB2-B17}

 \author{V.~E.~Stroud
         \inst{1,2,3}
          \and
J.~S.~Clark\inst{2}
\and 
I.~Negueruela\inst{4}
\and
P.~Roche\inst{1,2,3}
\and
A.~J.~Norton\inst{2}
\and
F.~Vilardell\inst{4}
           }

\institute{Faulkes Telescope Project, School of Physics and Astronomy,
 Cardiff University, Cardiff, CF24 3AA, United Kingdom 
            \email{vanessa.stroud@faulkes-telescope.com}
         \and
	    Department of Physics and Astronomy, The Open University,
 Walton Hall, Milton Keynes MK7 6AA, United Kingdom 
        \and
           Division of Earth, Space and Environment, University of
 Glamorgan, Pontypridd, CF37 1DL, United Kingdom
 \and
	   Departamento de F\'{i}sica, Ingenier\'{i}a de Sistemas y 
Teor\'{i}a de la Se\~{n}al, Universidad de Alicante, Apdo. 99, 03080
 Alicante, Spain
}

 \abstract {Massive, eclipsing, double-lined, spectroscopic binaries
 are not common but are necessary to understand the evolution of massive
 stars as they are the only direct way to determine stellar masses.
 They are also the progenitors of energetic phenomena such as X-ray 
binaries and $\gamma$-ray bursts.}
{We present a photometric and spectroscopic analysis of the candidate
 binary system Cyg OB2-B17 to show that it is indeed 
a massive evolved binary.}
{We utilise $V$ band and white-light photometry to obtain a light curve and
 period of the system, and spectra at different resolutions to calculate
 preliminary orbital parameters and spectral classes for the components.} 
{Our results suggest that B17 is an eclipsing, double-lined, 
spectroscopic binary with a period of $4.0217\pm0.0004$ days, with two 
massive evolved components with preliminary classifications of O7 and 
O9 supergiants. The radial velocity and light curves are consistent with
 a massive binary containing components with similar luminosities, and
 in turn with the preliminary spectral types and age of the association.}
{}

\keywords{binaries: eclipsing -- binaries: spectroscopic -- stars: early-type -- 
stars: fundamental parameters -- stars:evolution -- stars: individual: Cyg OB2 B17}

   \maketitle
%

\section{Introduction}
O stars are amongst the most massive and intrinsically luminous stellar 
objects found in galaxies. Since they are the only direct way to measure the masses and radii 
of stars, binaries are the perfect testbeds for studying the physical properties and evolution of 
such stars. Unfortunately, as reported by \cite{bon08}, less 
than 20 O stars have accurate ($\leq$10\%) dynamical mass estimates. 
Because of the scarcity of known massive, eclipsing, double-lined,
 spectroscopic binaries - and hence dynamical mass estimates for stars at
 different evolutionary states \citep[e.g.,][]{gie02} - the mass luminosity relation 
and theoretical evolutionary tracks of massive stars ($M\geq20M_\odot$) are
 currently poorly constrained by observations. 

In order to address this shortfall, much effort has been expended to identify 
further examples, utilising photometric and spectroscopic observations of
 young massive clusters such as the Arches \citep{mar08},
 Quintuplet \citep{fig99}, and Westerlund 1 (\citealt{cla05}; \citealt{rit09}). These observations 
indicate that the binary fraction is potentially very high\footnote{A large amount of 
observations and patience are necessary to determine the true binarity of a sample of
 stars in an open cluster as observed in \cite{san08}};
\cite{kob07} inferred it to be $\geq$ 80\% in Cygnus OB2 (Cyg OB2),
 while \cite{cla08} estimate the binary fraction of WR stars in
 Westerlund 1 to be $\geq$70\%. 
 Currently the
stars with the highest dynamical mass estimates are the twin WN6ha components 
of  WR20a within Westerlund 2 (83$M_\odot$+82$M_\odot$;
 \citealt{rau04}, \citealt{bon04}) and the newly discovered
 WN6ha binary NGC3603-A1 (116$\pm$31$M_\odot$ + 89$\pm$16$M_\odot$;
 \citealt{sch08}). This in turn has implications for the 
determination of the Initial Mass Function (IMF) for the clusters in question and, 
by extension, the empirically determined maximum mass possible for a 
star \citep[cf.][]{fig05}. Moreover, massive close binaries are the progenitors
 of such diverse energetic phenomena as supernovae, $\gamma$-ray bursts and
 X-ray binaries \citep{rib06}. Clearly the properties of the progenitor binary 
population must be known to constrain their formation channels.

Cyg OB2 is one of the most massive and richest associations in 
the Galaxy. It is $\sim$2 Myr old and 1.8 kpc away \citep{kim07}. 
Containing at least 60-70 O-type stars \citep{neg08}, its proximity and 
accessibility to optical studies have made it the focus of numerous 
observational campaigns to determine the properties of its massive stellar 
population \citep[e.g.][]{mas91, kno00, han03, com02, kim07, kim08}.

 Cyg OB2 B17 (\citealt{com02}; henceforth B17 and also known as V1827Cyg,
 2MASS J20302730+4113253, NSVS 5738756;
 $\alpha_{2000}$=20$^{h}$30$^{m}$27.3$^{s}$, 
$\delta_{2000}$=+41$^{o}13^{\prime}$25$^{\prime\prime}$; $V$=12.6) is a luminous,
variable member of the Cyg OB2 association. \citet{com02} observed the 
system as part of a near-infrared spectroscopic survey  and found the spectrum  
presented $Br$$\gamma$ emission, confirming it to be an evolved massive star. 
Follow up observations were made by \cite{neg08}, who found it 
to stand out from the rest of the members due to its variability and strong emission 
lines. They classified it as an Ofpe star and suggested it was a strong binary candidate.

This paper reports the first results of an extensive multi-epoch 
photometric and spectroscopic observational campaign on the binary candidate B17. 
Section~2 gives a description of 
the photometric and spectroscopic observations. Photometrically, the system was found 
to be variable and we report the analysis of the light curve in Section~3.
 More spectroscopic data were obtained permitting preliminary spectral and luminosity 
classification of the system; this analysis is shown in Section~4, 
along with descriptions of the long and short timescale variations of the spectra.
 The light and radial velocity curve modelling is presented in Section~5.
A discussion of the system, including its evolutionary status, is found 
in Section~6 and a summary is presented in Section~7.

Note that the central goals of this manuscript are to verify the binary hypothesis and 
present a preliminary spectral classification. A full analysis of the system,
 consisting of the deconvolution of an expanded 
spectral data set and subsequent model atmosphere analysis to 
determine the fundamental stellar parameters of the system will be presented in a 
future paper (Stroud et al. in prep).


\section{Observations and Data Reduction}
     
\subsection{Photometry}

The North Sky Variability Survey (NSVS; \citealt{woz04}) is a record of the 
sky at declinations higher than $\delta=-38\degr$ over the optical magnitude range 8 
to 15.5. It contains light curves of over 14 million objects. The data were taken between
 1999 April and 2000 March by the first-generation Robotic Optical Transient Search 
Experiment (ROTSE-I) at Los Alamos National Observatory, New Mexico. The telescope 
consisted of four unfiltered Canon 200mm telephoto lenses with f/1.8 focal ratio, 
each covering $8\fd2\times8\fd2$. These were equipped with AP10-cameras and  
Thomson TH7899M CCDs. The lenses had a typical point-spread function with a full width
 half maximum of $\sim$20$^{\prime\prime}$. In a median field, the bright unsaturated 
stars had a point-to-point photometric scatter of $\sim$0.02 mag and position errors 
within 2$^{\prime\prime}$. The calibrated images were passed through SExtractor software
 \citep{ber96}, reducing them to object lists. The data were accessed
 through the Sky Database for Objects in Time-Domain (SkyDOT) at Los Alamos National
 Laboratory. A total of 186 observations were obtained for B17.

 Additional $V$ band photometry was obtained by amateur astronomers Pedro Pastor Seva 
(observer 1) and Manuel M\'{e}ndez Marmolejo (observer 2) between 2007 April 03 and 2007 June 11.
Observatory 1 is located in Muchamiel (Alicante, Spain). The telescope used was an 8
 inch Vixen VISAC Schmidt-Cassegrain telescope. It has a field of view of
 24$^{\prime}$x20$^{\prime}$ and a focal ratio of f/8.The telescope was equipped with a SBIG ST10-XME 
CCD chip. Observatory 2 is located in Rota (C\'{a}diz, Spain). The telescope used was an 8 inch
 Meade LX200 Schmidt-Cassegrain telescope. It has a field of view of 16$^{\prime}$x12$^{\prime}$
(as a smaller SBIG ST7-XME chip was used) and a focal ratio of  f/6.3. Exposures varied between 3 and 4 
minutes giving a SNR  $\sim$200. The data were calibrated and analysed
 using Mira Pro\footnote{Mira Pro software is published by Mirametrics Inc., which has
 no connection with the Monterey Institute for Research Astronomy} and 
AIP4WINv2\footnote{AIP4WINv2 is published by Willmann-Bell, Inc.} packages. The apparent magnitudes were obtained
 using differential photometry with respect to a set of reference stars in the image with known magnitudes. 
In both cases the precission for the individual meassures is always better than 0.01 mag.

\subsection{Spectroscopy}
The spectra were obtained from several telescopes during the course of $\sim$three years. 
Table~\ref{obs} lists the full set of observations. The first set was observed with the 
1.52-m G. D. Cassini telescope at the Loiano Observatory (Italy) during the night
 of 2004 July 18. The telescope was equipped with the Bologna Faint Object 
Spectrograph and Camera (BFOSC) and an EEV camera. Grism 3 was used, which 
covers 3300-5800$\AA$ with a resolution of $\sim$6\AA.   
More spectra of the system were obtained with the 4.2-m William Herschel Telescope 
(WHT), in La Palma (Spain) equipped with the ISIS double-beam spectrograph, during
 service runs on 2006 June 11 and August 18. The instrument was fitted with the R300R grating
 and MARCONI2 CCD in the red arm and the R300B grating and EEV12 CCD in the blue
 arm. Both configurations result in a nominal dispersion of 0.85\AA/pixel (the
 resolution element is approximately 3 pixels in the blue and 2 pixels in the red). 

The system was also observed with the 2.5m Isaac Newton Telescope (INT) in La
 Palma on 2006 September 10-11. The telescope was 
equipped with the Intermediate Dispersion Spectrograph (IDS), fitted with a R632V
 grating and EEV10 CCD.
Another 12 spectra were obtained during a dedicated run on 2007 August 21-22 at the WHT.
 The system was observed in the blue arm with grating R1200B
 (nominal dispersion of $\sim$0.23\AA/pixel). 
All the spectra were reduced with the {\em Starlink} packages {\sc ccdpack} \citep{dra00}
 and \textsc{figaro} \citep{sho97} and analysed using {\sc figaro} and {\sc dipso} \citep{how98}.


\begin{table}    
\caption {Log of Spectroscopic Observations. } 
\label{obs}
    \centering
         \begin{tabular}{c c c c c c c}

\hline\hline						
       Date    &Telescope      &Nominal       &Wavelength        &Phase\\
               &  /Instrument  & Dispersion   & Range (\AA)        &  \\
               &               & (\AA/pix)      &                  &\\
\hline 
04/07/18      &Cassini/BFOSC       & 6&3800-6400&0.437
\\
04/07/18      &Cassini/BFOSC      &	6&6100-8200&0.450
\\
06/06/11    &WHT/ISISB    & 0.86 &3200-5200 & 0.036
\\
06/06/11    &WHT/ISISR    & 0.93 &5400-8000 & 0.036
 \\
06/08/18    &WHT/ISISB    & 0.86 &3200-5200 & 0.929
 \\
06/08/18    &WHT/ISISR    & 0.93 &5400-8100 & 0.931
 \\
06/09/10    &INT/IDS      & 0.9 &3900-5000 & 0.600
 \\
06/09/11    &INT/IDS      & 0.9 &3900-5000 & 0.840
 \\
06/09/11    &INT/IDS      & 0.9 &3900-5000 & 0.891
 \\
07/07/21    &WHT/ISISB    & 0.86  &3600-5300 & 0.738
 \\
07/07/21    &WHT/ISISR    & 0.93 &5400-8300 & 0.738
 \\
07/07/21    &WHT/ISISR    & 0.93 &5400-8300 & 0.740
 \\
07/07/21    &WHT/ISISR    & 0.93 &5400-8300 & 0.741
 \\
07/08/21    &WHT/ISISB    & 0.23 &3900-4750 & 0.380
 \\
07/08/21  &WHT/ISISB    & 0.23 &3900-4750 & 0.385
 \\
07/08/21  &WHT/ISISB    & 0.23 &3900-4750 & 0.397
 \\
07/08/21  &WHT/ISISB    & 0.23 &3900-4750 & 0.416
 \\
07/08/21  &WHT/ISISB    & 0.23 &3900-4750 & 0.436
 \\
07/08/21  &WHT/ISISB    & 0.23 &3900-4750 & 0.453
 \\
07/08/22  &WHT/ISISB    & 0.23 &3900-4750 & 0.629
 \\
07/08/22  &WHT/ISISB    & 0.23 &3900-4750 & 0.634
 \\
07/08/22  &WHT/ISISB    & 0.23 &3900-4750 & 0.648
 \\
07/08/22  &WHT/ISISB    & 0.23 &3900-4750 & 0.664
 \\
07/08/22 &WHT/ISISB    & 0.23 &3900-4750 & 0.683
 \\
07/08/22  &WHT/ISISB    & 0.23 &3900-4750 & 0.701
 \\

\hline			
       
        \end{tabular}
   
\end{table}


\section{Photometry}

The {\em Starlink} software {\sc period} \citep{dhi01} was used on both the 
NSVS and amateur photometry to search for a modulation period in the photometric data,
 using phase dispersion minimisation, $\chi^{2}$ of sine fit vs frequency and 
string-length vs frequency methods. The resultant periods were consistent and favoured
 an orbital period of $4.0217\pm0.0004$ days
(the error being estimated from the spread in the periods derived from the different
 methods employed). When both data sets were folded together they were found to be 
in phase; therefore, given that there appears to be no change of shape or shift in 
period in the 7 years between the NSVS and amateur observations, we are confident 
that the period determined is accurate to within the errors quoted.


\begin{figure}[h!]
   \centering
        \resizebox{\columnwidth}{!}{\includegraphics[trim = 0mm 0mm 5mm 60mm, clip]{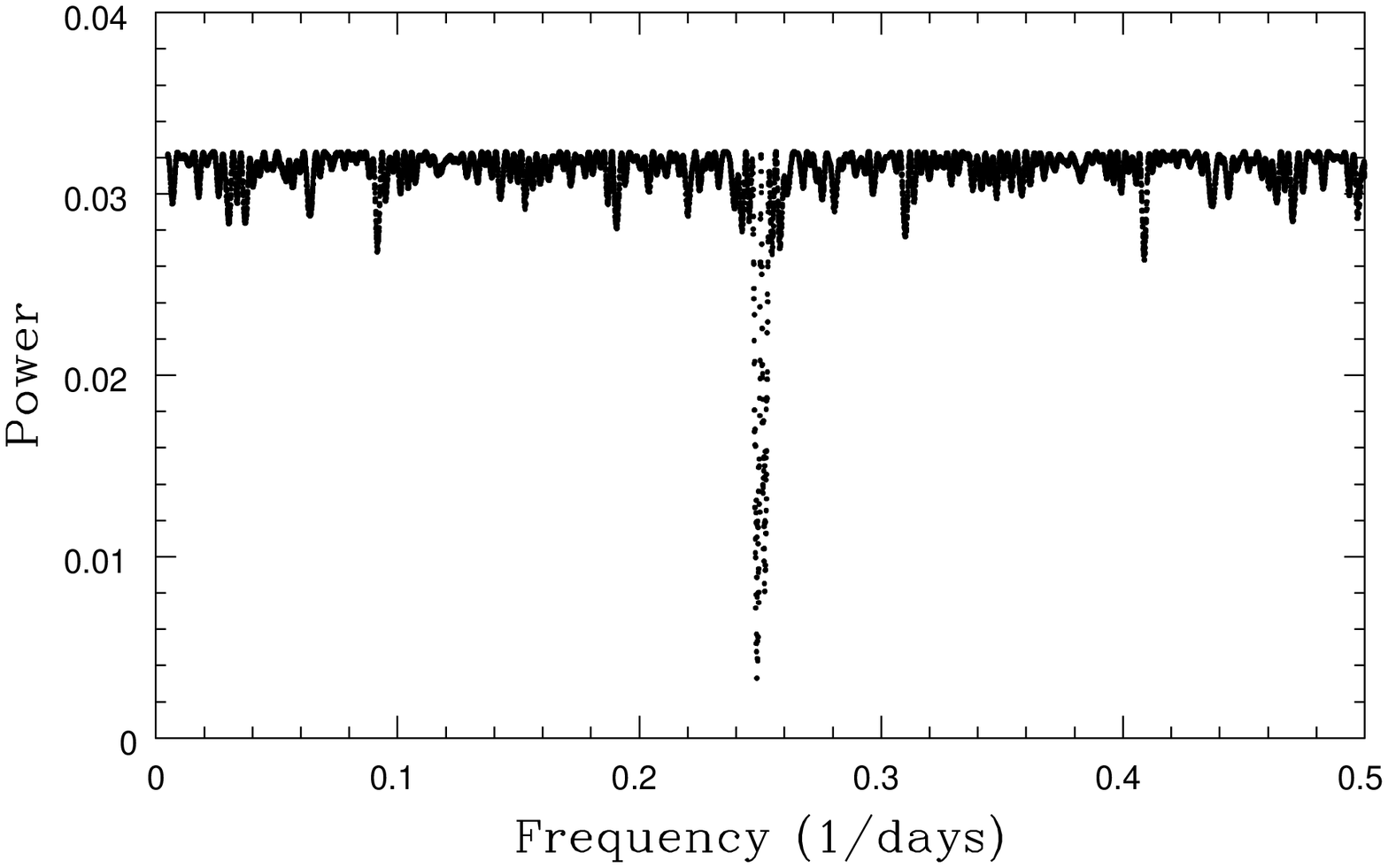}}
     \resizebox{\columnwidth}{!}{\includegraphics[trim = 0mm 0mm 5mm 60mm, clip]{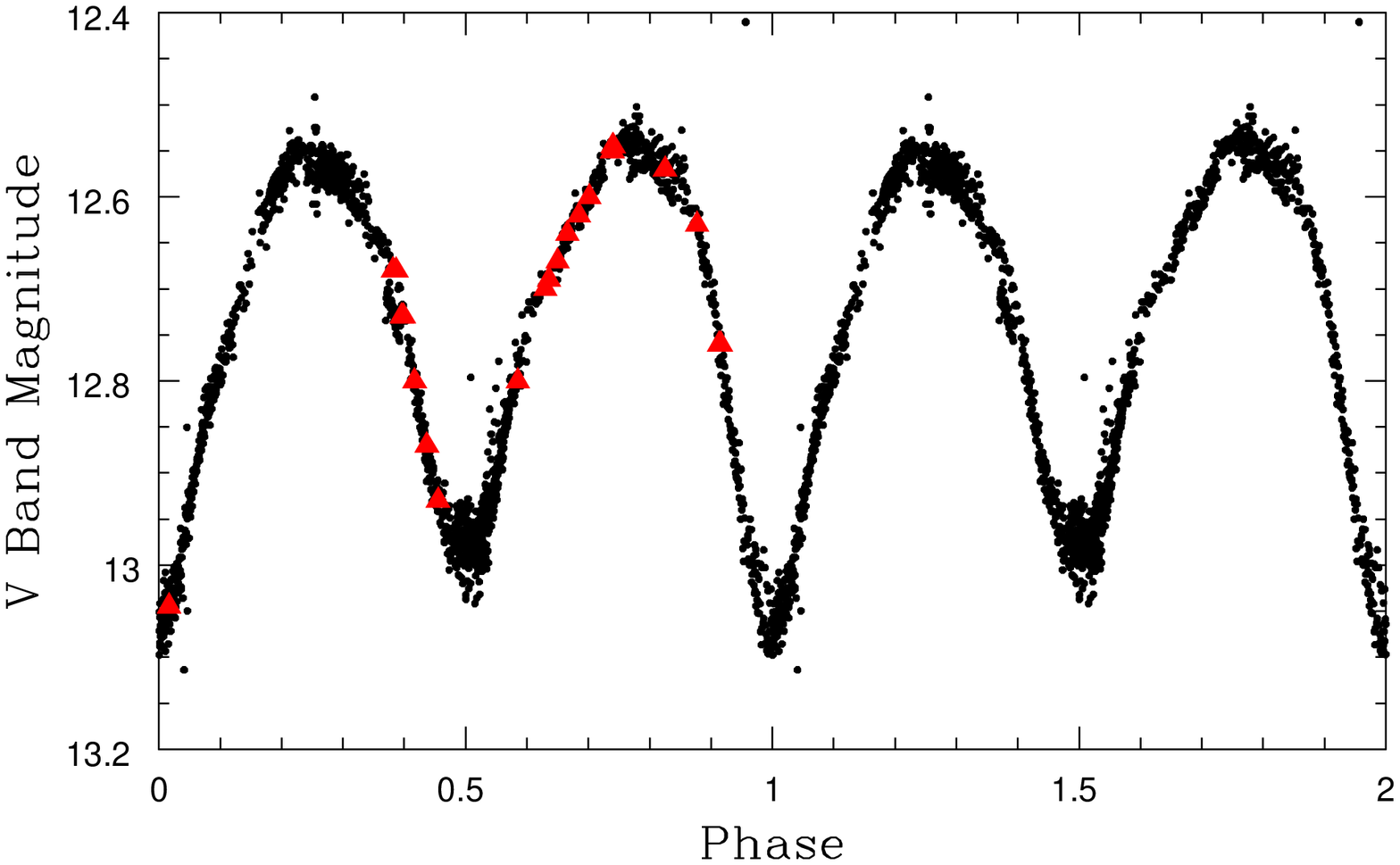}}
   \caption{Upper panel: Periodogram for B17 data using the reduced-$\chi^2$ technique.
 Lower panel: Light curve folded on a 4.0217 day period. The red triangles show the phases for which spectra were obtained.} 
 \label{fig:periodogram}
\end{figure}

Fig.~\ref{fig:periodogram} shows the periodogram obtained using Period with the reduced-$\chi^2$ 
technique (top) and the 
$V$ band light curve folded on a 4.0217 day period, along with the phases of the spectra obtained (bottom).

 The shape of the light curve 
suggests that the system is a semi-detached binary; Both  minima 
are narrow and demonstrate different eclipse depths, although the $0.5-0.6$~mag range indicates the 
almost complete eclipse of each star 
and hence that the two stars are of similar size but with 
different luminosities (in a contact system, the temperature of both stars should be the same)
 There appears to be an asymmetry on the light curve between
 phases $\phi=0.6 - 0.75$ which is observed in both sets of data. \cite{hil05} found similar depressions 
in eclipsing binaries in the Small Magellanic Cloud and
 \cite{bon08} found a similar depression in the light curve for the binary LMC-SC1-105. They attributed
 these asymmetries to the presence of a mass-transfer stream. \cite{lin09} also observed this asymmetry 
in the overcontact binary Cyg OB2 $\#$5 and attributed this to the secondary being brighter on its leading
 side due to the colliding wind region; first discussed by \cite{rau99}. 

The quadratures display an O'Connell effect - where the maxima are of different brightnesses -  of $\sim$0.02 mag which
is likely due to variations in the brightness of the stellar surface(s) due to mass transfer between the components.

 The principal minimum was estimated using the amateur data to be at $JD=2,454,272.527\pm0.005$.
  From the light curve, the ephemeris for the primary eclipse was found to be:
\begin{equation}
minI~=~2454272.527~+~4.0217E~~{\rm (JD)}         
\end{equation}

\noindent where $E$ is the number of orbital cycles after the given epoch.

 We note that our period determination is in agreement with a value recently 
established by \cite{ote08} using the same data from NSVS as used in this paper.

\section{Spectroscopy}

The spectrum of B17 is found to be highly variable on both short ($<$day) and long
 ($\sim$year) timescales, with the most prominent features in the spectra being
 H, \ion{He}{i}, \ion{He}{ii} and \ion{N}{iii} lines in absorption and \ion{He}{ii}
 and \ion{N}{iii} lines in emission. Wavelength shifts are observed 
for different lines, some of which become double in some spectra suggesting that it
 is a binary system with at least one hot component, as revealed 
by the presence of \ion{He}{ii} and \ion{N}{iii}.

\subsection{Long Term Variability}
 Fig.~2 shows blue-violet spectra (4000-4900\AA) obtained at random phases between 2004 July and 2007 August.
 The most prominent elements observed are H, \ion{He}{i}, 
\ion {He}{ii} and \ion{N}{iii} in absorption and \ion{He}{ii}~$\lambda$4686 and 
\ion{N}{iii}~$\lambda$$\lambda$4634-40-42 in emission. The spectra show strong, highly 
variable \ion{N}{iii} and \ion{He}{ii} emission lines; in 2004 July 18 these are of
 a similar strength but by 2006 June 11 the \ion{He}{ii} line is almost three times
 the intensity of the \ion{N}{iii} lines. The \ion{N}{iii} profile is also variable;
 it appears as two separate peaks in some of the spectra (2006 June 11,
2006 September 10-11, 2007 July 21) and it is almost completely blended in the
 spectra taken on 2004 July 18 and 2006 August 18, although this in part could be due
 to the low resolution of the spectrum from 2004 July 18.

The hydrogen lines are also highly variable; this is particularly evident for
 H$\gamma$ when compared to the diffuse interstellar band (DIB) at $\lambda$4428. 
H$\beta$ shows a P-Cygni profile varying in both width and strength suggesting the 
stellar wind is highly variable.
 Finally, the \ion{He}{i}~$\lambda$4471/ \ion{He}{ii}~$\lambda$4541 ratio, which is
 employed as a standard diagnostic for the spectral type of early stars 
\citep{wal90}, is also observed to vary over the course of the 
observations. 


\begin{figure}[h]
   \centering
  \label{fig:bluespec}
   \resizebox{\columnwidth}{!}{\includegraphics[trim = 0mm 0mm 40mm 0mm, clip]{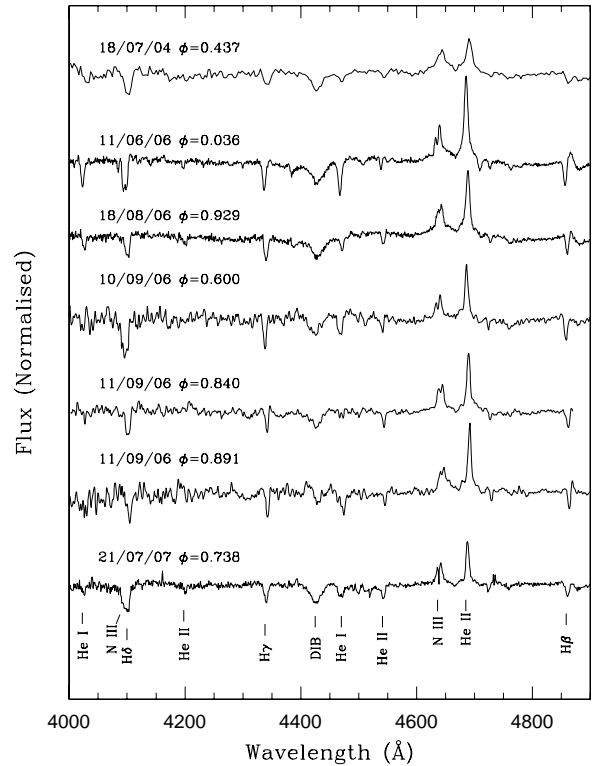}}
   \caption{Blue-violet spectra of B17 obtained at different phases over a 
three year period, with the most prominent lines labeled.}
   \label{blue}
\end{figure}

The most prominent line in the yellow spectra (5600-6000\AA) is the varying P-Cygni
 profile in \ion{He}{i}~$\lambda$5875, which blends with the \ion{Na}{i}
 interstellar lines (Fig.~3). When compared to the nearby DIBs at 5780-5800$\AA$, 
the absorption trough in \ion{He}{i} line was observed to vary by a factor of 4 in
 intensity relative to the nearby DIB features, with the emission
 component also varying in strength by a factor of 2. Other lines in the yellow
 spectra include \ion{C}{iii}~$\lambda$5696 and \ion{C}{iv} around $\lambda$5810. 


\begin{figure}[!ht]
   \centering
 \label{fig:greenspec}
     \resizebox{\columnwidth}{!}{\includegraphics[trim = 0mm 0mm 40mm 0mm, clip]{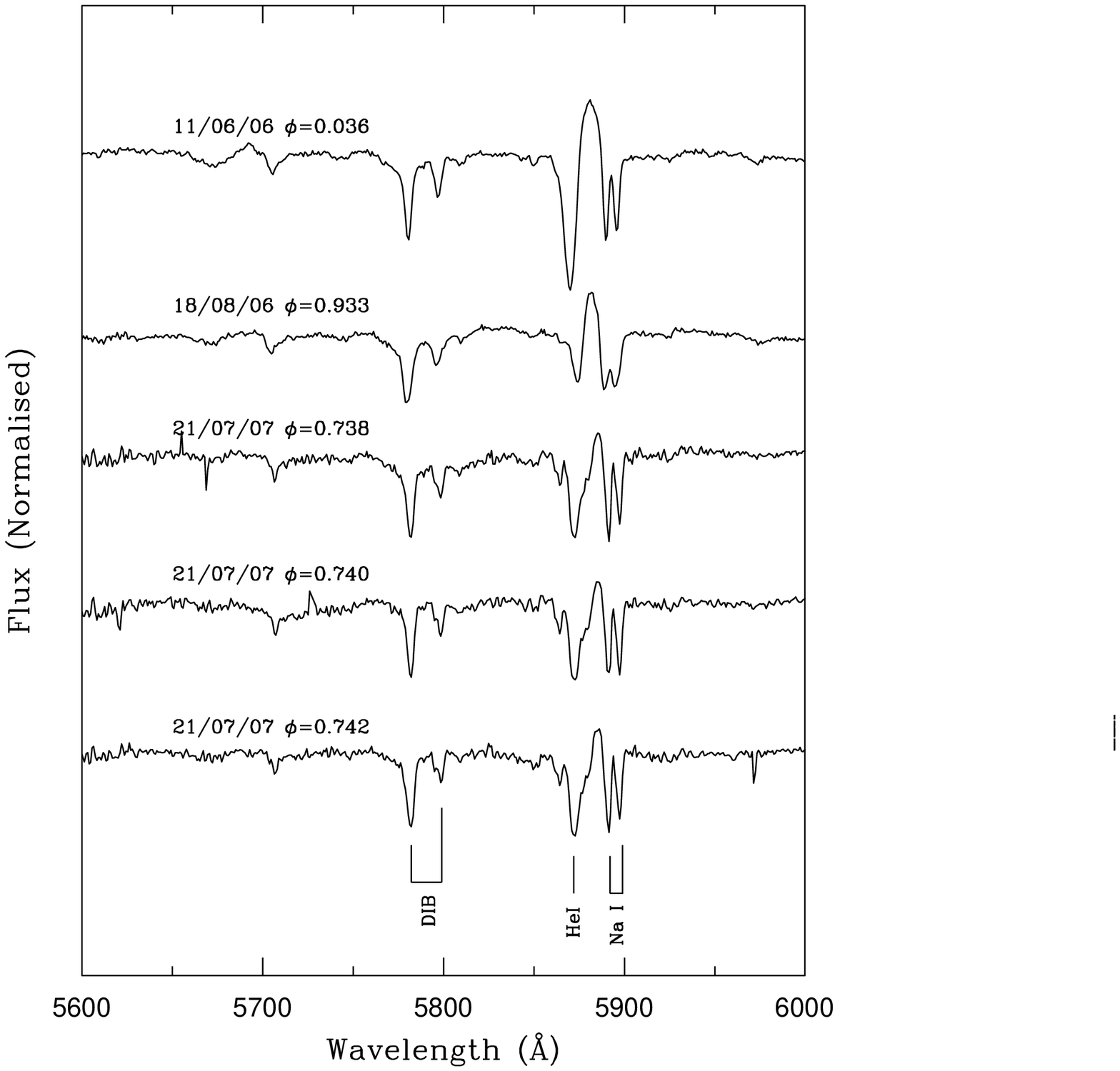}}
   \caption{B17 yellow spectra showing variability at different phases taken over
 a period of three years. Prominent lines are labeled.}
\end{figure}

The most prominent feature in the red spectra (6350-6650\AA) is the H$\alpha$ line 
which also demonstrates a highly variable P-Cygni profile  which is blended with 
\ion{He}{ii} absorption features in all the red spectra, complicating analysis of 
the profile. It has a double peak profile which varies in intensities with the redder
 peak being between 2 and 4 times stronger than the bluer peak. The depth of the 
\ion{He}{ii} line, which presumably contributes to the double peaked morphology 
 is also variable (Fig.~4). Preliminary calculations for the velocities of the blue 
edge of the H$\alpha$ profile ($v_{{\rm edge}}$) have a range of 2260 - 2625 kms$^{-1}$
 with an average 2440$\pm$50 kms$^{-1}$. The $v_{{\rm edge}}$ is related to the
 terminal velocity of the wind; in the extreme ultraviolet, the terminal velocities
 for OB stars are 15\%-20\% smaller than the edge velocities \citep{pri90}. Variations 
in the wind profiles of OB stars are thought to be a consequence of highly structured
 winds on multiple scales, and while it appears likely that the variations observed 
for B17 arise due to a highly anisotropic circumstellar envelope observed at differing
 lines of sight throughout the orbital period, the current limited data set offers
 little prospect of a more explicit physical interpretation.


\begin{figure}[!ht]
   \centering  
 \label{fig:redspec}
      \resizebox{\columnwidth}{!}{\includegraphics[trim = 0mm 0mm 40mm 0mm, clip]{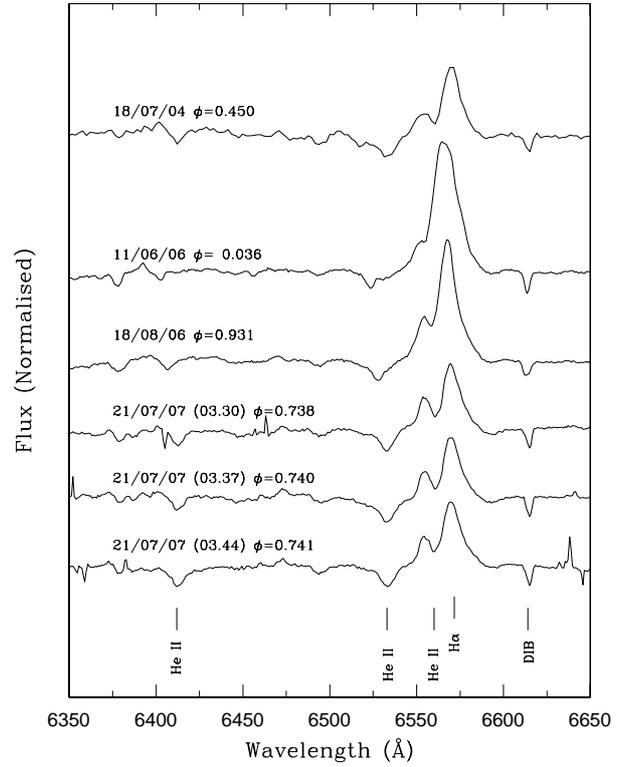}}
   \caption{B17 red spectra showing H$\alpha$ variability at different phases 
taken over a period of three years. Prominent lines are labeled.}
\end{figure}

Unfortunately, no spectra with the same orbital phases have been obtained at 
different epochs and so it has not been possible to search for unambiguous long term 
secular variability in the current data set.

In Fig. 5 we plot the EW of the \ion{He}{ii}~$\lambda$4686 line against orbital phase, finding the EW to
 be broadly anticorrelated with the photometric light curve, increasing by a factor of 
two during the entry to eclipse. This result suggests that the strength of the line
 remained $\sim$constant over the segments of the orbital period sampled by the
 observations, implying that the change in EW is primarily due to dilution by the
 variable continuum.


\begin{figure}[!ht]
   \centering  
 \label{fig:EWs}
         \resizebox{\columnwidth}{!}{\includegraphics[trim = 0mm 0mm 0mm 0mm, clip]{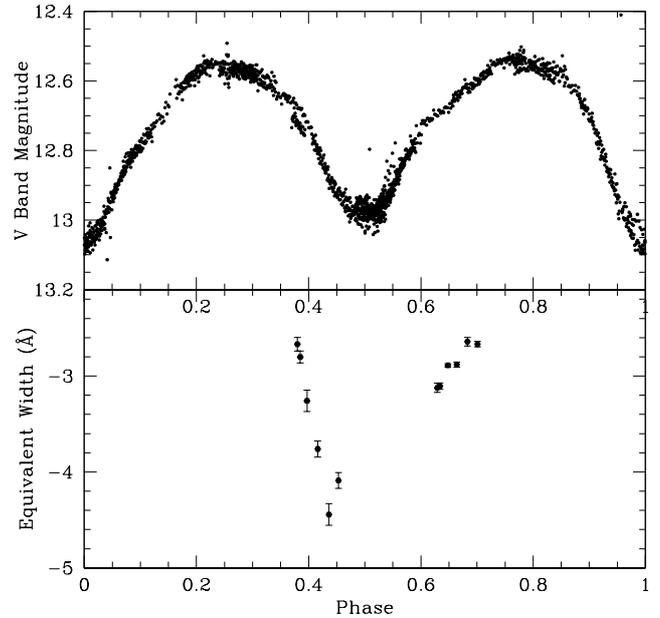}}
   \caption{Upper panel: Light curve folded on a 4.0217 day period. Lower panel: Equivalent widths of the \ion{He}{ii}~$\lambda$4686 
emission line for the WHT data taken in 2007 August plotted on the same orbital period.}
\end{figure}

\subsection{Short Term Variability}

The set of 12 high resolution spectra obtained during the nights of the 2007 August 
21-22 cover the phases $\phi$=0.38-0.453 and $\phi$=0.629-0.701 of one period. The 
spectra obtained during the first night (Fig.~6 upper panel) have narrower H and \ion{He}{ii}
 absorption lines lines compared to the spectra from the second night. During the first
 night, the \ion{He}{i}~$\lambda$4471 /\ion{He}{ii}~$\lambda$4541 ratio - used for 
classifying O stars - changes monotonically in the sense of increasing \ion{He}{i}~$\lambda$4471 
strength, reaching unity in the final spectrum.

 A second, blueshifted absorption component appears on the shoulder of the 
\ion{He}{i}~$\lambda$4471 line, becoming stronger with time; similar evolution is also 
observed in the  H$\delta$ profile. Initially, the \ion{N}{iii}~$\lambda$4634-40-42  
emission is highly blended, with the redshifted peak being the strongest. These peaks 
start to separate on the last spectrum of that night ($\phi$=0.453). The \ion{He}{ii}~$\lambda$4686
 emission feature appears to be non-symmetric and strengthens throughout
 the night. The \ion{He}{i}~$\lambda 4026$, \ion{He}{ii}~$\lambda$4200 and H$\gamma$
 lines do not vary significantly throughout the first night.

During the 17hrs between the last observation of night one ($\phi$=0.453) and the
 first of night two ($\phi$=0.629) the spectrum has clearly evolved. The 
\ion{He}{i}~$\lambda$4026, \ion{He}{ii}~$\lambda$4200 and H$\gamma$ lines are broader
 and non symmetrical. On the second night (Fig.~6 lower panel), 
the \ion{He}{i}~$\lambda$4471 / \ion{He}{ii}~$\lambda$4541 ratio is $>$ 1 on the first spectrum 
of the second night and decreases until it reaches unity on the last spectrum 
($\phi$=0.701). Significant changes are apparent in the \ion{He}{i}~$\lambda$4471 
absorption profile, which is much broader and appears to be composed of two troughs,
 with the bluer trough being stronger. An additional, distinct weak blueshifted 
absorption line is also present  and  becomes stronger with time. The H$\delta$ line
 appears to be the result of the blending of two lines, showing a wide profile with 
two troughs. The bluer trough is stronger than the redder during the second night.
 The \ion{N}{iii}~$\lambda$4634-40-42 emission is observed as two separate peaks,
 with the redder being the strongest and the intensity of both increasing through 
the night. The \ion{He}{ii}~$\lambda$4686 emission feature appears to be symmetric 
compared to the first night and weakens throughout the night. We note that (as
 mentioned in Section 3) the light curve demonstrates an asymmetry - possibly due to
 mass transfer - between phases $\phi \sim$0.6--0.75. Given this is coincident with
 the  observations during night two, it is tempting to attribute the spectral changes
 (in part) to such a process.


\begin{figure*}[h!]
   \centering
 \label{fig:blueshort}
   \includegraphics[trim = 0mm 8mm 0mm 0mm, clip,scale=0.55, angle=270]{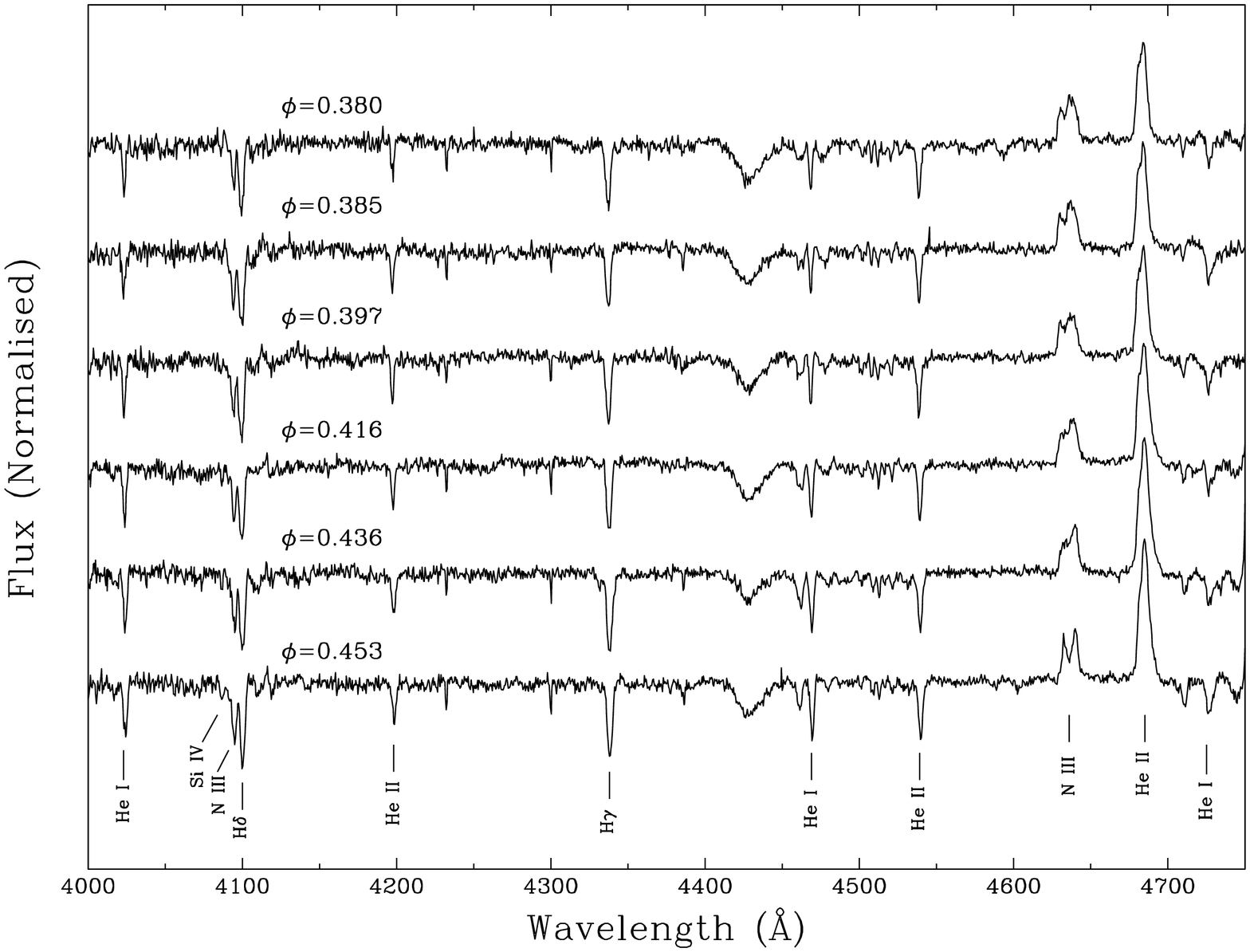}
   \includegraphics[trim = 0mm 8mm 0mm 0mm, clip,scale=0.55, angle=270]{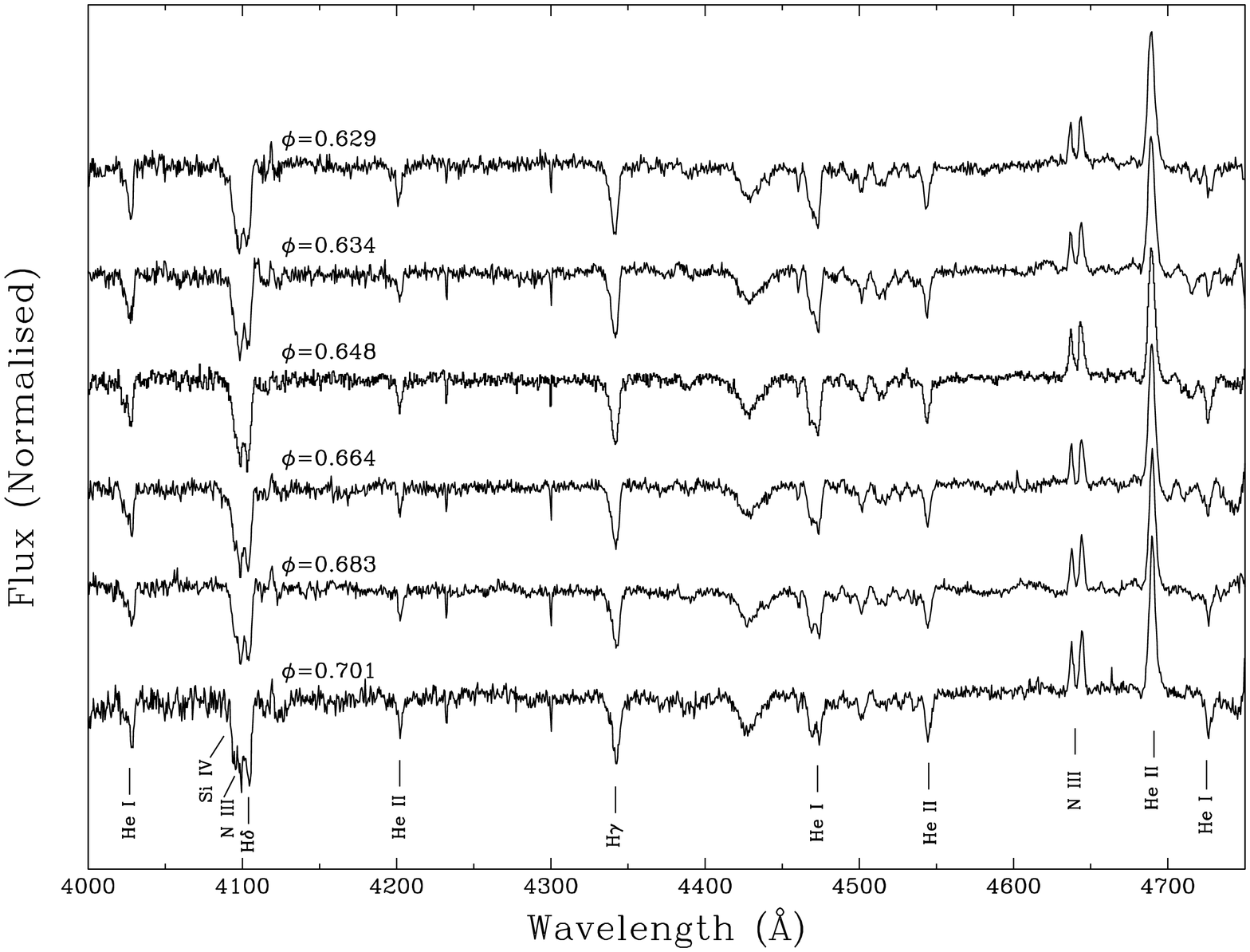}
   \caption{Blue spectra obtained with the WHT on the night of 2007 August 21 and 22. 
The main transition lines are shown. On the regions where the lines are blended,
 accuracy of the identified lines is not certain and will be clarified with spectra
 dissentangling.}
\end{figure*}

\subsection{Spectral Classification}

Given that the current spectroscopic dataset does not span an entire orbital cycle
 we have not attempted a formal deconvolution of the blended spectra. Instead we 
have simply used the spectra closest to the primary and secondary eclipses to 
perform a {\em preliminary} classification of the components of the system,
using the stellar atlas of \cite{wal90} and \cite{wal00}.

The spectrum at $\phi$=0.453 shows a \ion{He}{ii}~$\lambda$4541 / 
\ion{He}{i}~$\lambda$4471 absorption-line ratio of one, suggesting a likely spectral
 classification of O7. Both \ion{He}{ii}~$\lambda$4686 and \ion{N}{iii}~$\lambda\lambda$4634-40-42 
are in strong emission which is characteristic of an
 Of supergiant, implying an initial classification of O7~Iaf. Fig.~7 upper panel shows this 
spectrum along with comparison spectra. The spectrum lacks the \ion{Si}{iv}~$\lambda$4089 
line, which is unexpected for an O7 SG.\footnote{We note that while this line is 
not present in the spectrum of the O7 supergiant Sanduleak 80, given the reduced
 metalicity appropriate for a SMC star it is not clear it provides a valid comparison.}
\cite{wal01} mentions that a similar effect is found in spectra in the Small Magellanic
 Cloud where no \ion{Si}{iv}~$\lambda$4089 is observed and he attributes this to the absorption
 and emission features cancelling each other. Therefore we suggest a preliminary spectral
 classification of O7~Iaf+ for this component.

\begin{figure}[!ht]
   \centering
\label{fig:specclass}
 \resizebox{\columnwidth}{!}{\includegraphics[trim = 10mm 9mm 5mm 60mm, clip]{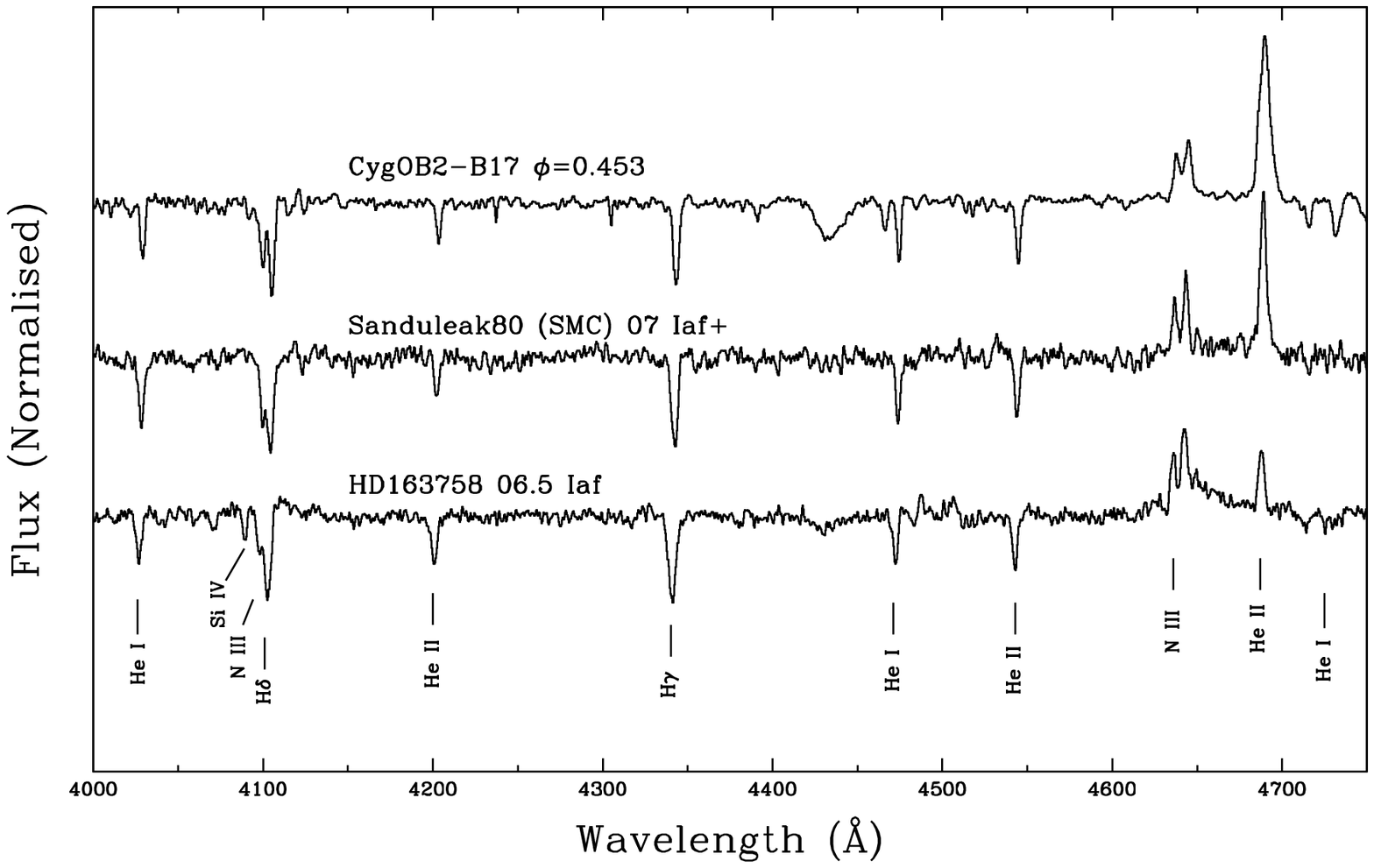}}
 \resizebox{\columnwidth}{!}{\includegraphics[trim = 8mm 0mm 5mm 75mm, clip]{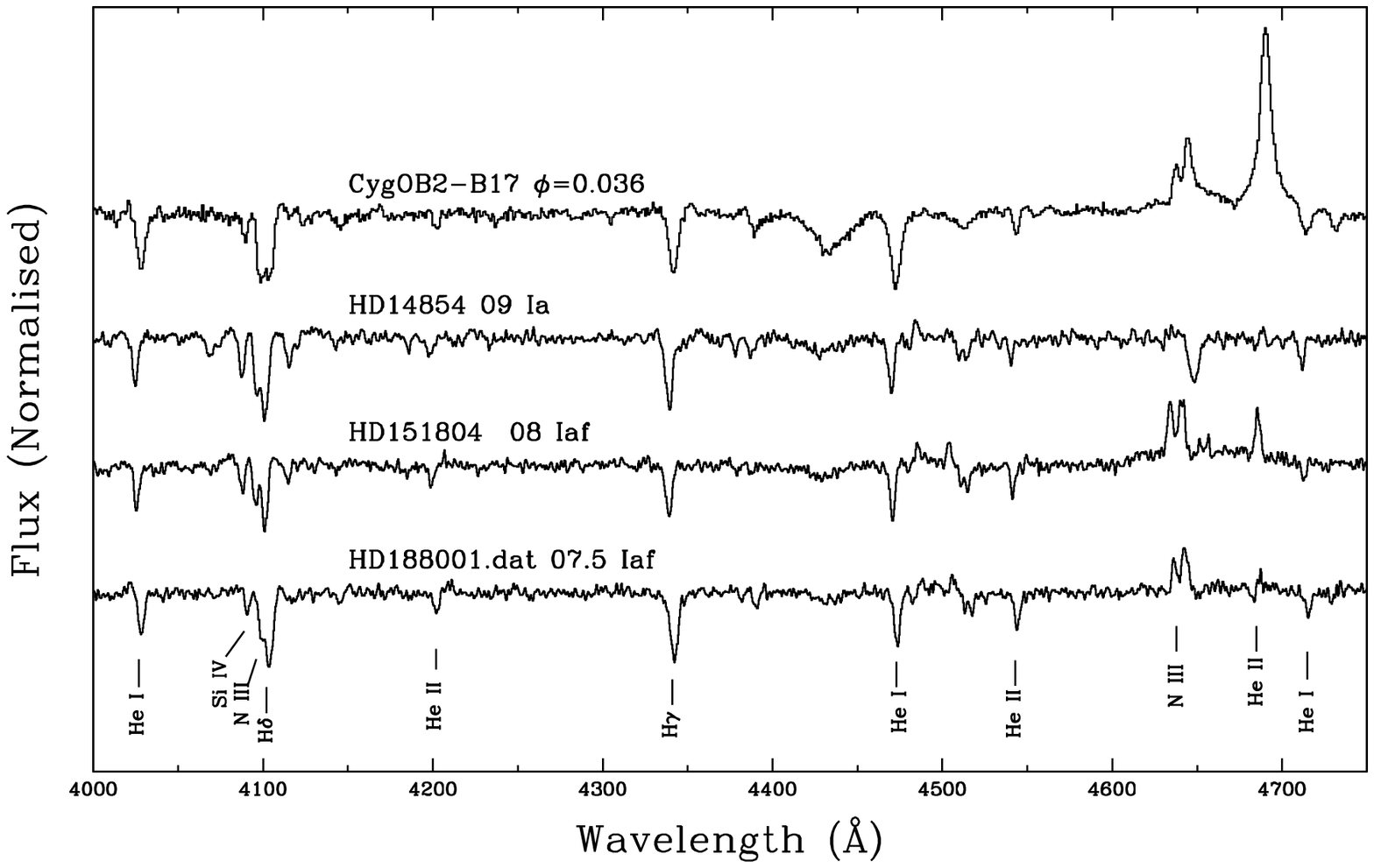}}

   \caption{Upper panel: B17 spectrum obtained with the WHT on the night of 2007 August 21,
 close to the secondary eclipse, compared with Of supergiant spectra from the
 Digital Atlas of Stellar Classification \citep{wal90}.
 The spectrum is most compatible with an O7Iaf classification. Double lines are observed
 even though the system is at eclipse which we currently cannot account for and will be
 further studied with the spectra dissentangling. Lower panel: B17 spectrum obtained 
with the WHT on the night of 2006 June 11,
 close to the primary eclipse, compared with Of supergiant spectra from the Digital
 Atlas of Stellar Classification \citep{wal90}. The spectrum is
 more comparable to the O9 classification.
}
\end{figure}

The spectrum at $\phi$=0.036 (Fig.~7, lower panel) shows a \ion{He}{ii} $\lambda$4541 / 
\ion{He}{i} $\lambda$4471 absorption-line ratio smaller than one, indicating a 
spectral type later than O7. The \ion{He}{ii} $\lambda$4686 and \ion{N}{iii}
 $\lambda$$\lambda$4634-40-42 are also in strong emission, implying an Of 
supergiant. The spectrum also presents a stronger Si \textsc{iv} 4089 absorption
 line compared to the spectrum at $\phi$=0.453 although it is still weaker than
 expected for a late O star. Therefore we suggest a preliminary spectral
 classification of O9~Iaf for this component.

 Since the combination of spectra of the preliminary
 classifications of an O7 and an O9 components would not give a HeI4471/HeII4541 ratio less
 than unity (see Section 4.2), it suggests there is emission in-filling of HeI4471.

\subsection{Radial Velocities}

The high resolution blue spectra show systematic night to night variations in the
 H, He and N lines, revealing significant radial velocity shifts. Radial velocities 
were determined for the primary lines which appear not to be blended
 (\ion{He}{ii}~$\lambda$4200, \ion{He}{ii}~$\lambda$4541, H$\gamma$~$\lambda$4340) in the high resolution 
spectra obtained on the WHT observing run in 2007 August. Gaussians were fitted to the line profiles using 
the {\sc dipso} emission line fitting command ({\sc elf}; the results are discussed in Sec.~5). The lines for the
 secondary appear to be blended and were not measured. The radial velocities for the secondary will be
 measured in the future paper after the spectra have been dissentangled.

\section{Light and radial velocity curve analysis}

In order to reproduce the observed characteristics of the photometric light curve, we analysed it with 
the 2003 version of the \citet[W-D;][]{wil71} code. 
For the analysis, the ROTSE (with an effective wavelength similar to Johnson $R$)
and amateur $V$ light curves were modeled as different datasets. In all cases,
detailed reflection-model and proximity-effect corrections were included.
Considering the spectroscopic analysis, the temperature of the primary ($T^{\rm
P}_{\rm eff}$) was fixed to 35\,000~K, and the bolometric albedo and gravity
brightening coefficients were set to unity, as generally found for stars with
radiative envelopes. In addition, a circular orbit was adopted, as suggested by
the equal separation between both (primary and secondary) eclipses, and a
rotation rate synchronized with the orbital period was assumed for both
components. The fitting process was carried out iteratively until three
consecutive solutions provided differential corrections for all the parameters
smaller than twice their internal errors. 

Considering the characteristics in the light curves described above, the first
runs in the modelling with W-D assumed a {\em semi-detached} configuration.
Numerous attempts were performed with several mass ratios and with either the
primary or the secondary component filling the Roche lobe.  However, {\em in all cases,
 the fits provided solutions where both stars tended to be in contact}.
Therefore, despite the properties of the lightcurve suggesting a semi-contact configuration
we finally attempted an over-contact solution - the results of which are described 
here - where each component could have a different temperature, as observed
from the different depths of the eclipses. For this model, the mass ratio was
set  to $q=0.75$ - as might be expected for O7 and  O9 super-giant components - noting that 
the mass ratio cannot be  smaller than ~0.4\footnote{For a mass ratio of 0.4, the implied mass of the primary 
  -considering the minimum amplitude of the radial velocity curve-  is over 120 M$_{\odot}$, regardless of the 
light curve analysis.}. The time of minimum ($t_{\rm min}$), the period ($P$), the orbital inclination
($i$), the temperature of the secondary component ($T^{\rm S}_{\rm eff}$), the
surface potential ($\Omega^{\rm P}$) and the luminosity of the primary
component ($L^{\rm P}$) were all left as free parameters in the fit. 

The $V$ light curve (which has smaller photometric errors) clearly reveals that one quadrature is
$\sim$0.02 mag brighter than the other. Two main solutions were attempted to
explain the observed O'Connell effect, one with an equatorial hot spot (30\%
hotter than the photosphere) on the primary component and another one with the
spot being on the secondary component. The size and position of the spot were
left as free parameters and also fitted at each run, although we emphasise  that the 
size and temperature of the spot are strongly correlated. The fits with the spot on
the primary component were finally adopted, since they provided slightly
smaller errors. In addition, the position of the spot (oriented roughly towards
the secondary component) can more easily be explained as the interaction of the
stellar winds.

Together with the light curve analysis, the radial velocities were also used to
constrain the solution. Radial
velocities were fitted separately from the light curve to avoid the larger
number of photometric observations dominate in the solution; the  parameters fitted 
being the semi-major axis ($a$) and the systemic velocity
($\gamma$). The rms of the fit is 9.5 km~s$^{-1}$ for the radial
velocities, 0.023 mag for the $V$ light curve and 0.048 mag for the ROTSE light
curve. Unfortunately, the accuracy of the solution obtained is strongly dependent on the 
mass ratio adopted. Nevertheless, Fig.~8 and Fig.~9 
show the light curve and radial velocity curve with the best fit and Table \ref{lcparam}
 gives the parameters obtained from this solution.

Despite our efforts we are still not completely satisfied with this fit. In particular, we
 are still unable to adequately fit the egress from secondary minimum  nor the 
subsequent lightcurve between phases 0.6-0.9. Moreover it is expected that the temperatures 
of stars in over-contact binaries should be $\sim$equal where a ratio of 0.85 was found between
 the secondary and primary. As such we regard the parameters presented in Table 2 as provisional 
at present. We anticipate that  the determination of radial velocities for the secondary
component, which will directly constrain the binary mass ratio
 will greatly clarify the fundamental properties of the components and
the configuration of this EB system.

 Finally, the lightcurve modelling, allows us to address the distance to B17, and by extention
the Cyg OB2 association. Adopting the bolometric
 corrections from \cite{mar05} we may use the bolometric luminosities determined above (See Table 2) to
calculate the absolute $V$ magnitudes of both components.
The $V$ band reddening was then calculated by following
 a similar procedure used by \citealt{neg08} (see Section~6)
With the absolute $V$ magnitudes, and the $V$ band reddening for B17, along with the $V$ band values for 
the two minima in the lightcurve, the distance modulus was calculated to be 10.9-11.3 which
 corresponds to a distance of about 1.5-1.8~kpc. While we regard these values as provisional due to
 the difficulties in modelling described above, these are consistent with the 
commonly adopted distance estimate of 1.7~kpc. (e.g. \citealt{han03}, \citealt{tor91} and  \citealt{kim07}). Therefore they do
 not agree with the distance estimate of 900-950~pc obtained by \cite{lin09} from 
the lightcurve modelling of Cyg OB2 $\#$5.

A distance of 900~pc would imply radii that are $\sim$half those currently derived from the modelling.
 The derived radii scale linearly with the semi-major axis and the semi-major axis in turn is strongly
dependant on the assumed mass ratio. For this system, a semi-major axis a factor of 2 smaller would
 imply a mass ratio of 6. We consider this to be extremely unlikely, but not impossible.


\begin{table}    
\caption {Results from the analysis of the light and radial velocity curves: The errors shown 
should be considered internal errors of the fit. Any possible systematic errors
 are not included.}

\label{lcparam}
    \centering
    
        \begin{tabular}{l c c }

\hline\hline						
Parameter                                 &   Value\\
\hline 

T$_{0}$ (MJD)                             & 4272.534 $\pm$ 0.004
\\          
Period, $P$                               & 4.02174 $\pm$ 0.00003 days
\\
Inclination, $i$                          & 72 $\pm$ 1.5
\\
Eccentricity, $e$                         & 0 (Fixed)
\\
Mass ratio $q$                            & 0.75 (Fixed)
\\
Temperature ratio                         & 0.85 $\pm$  0.02
\\
Semi-major axis ($R_{\odot}$)             & 50 $\pm$ 1
\\
Systemic velocity                        & -43 $\pm$ 4 km s$^{-1}$
\\
\hline 
\\
    
\hline\hline						
Parameter                                 &   Primary              &    Secondary\\
\hline 
Mass ($M_{\odot}$)                        &   60 $\pm$ 5          &     45 $\pm$ 4
\\
Radius ($R_{\odot}$)                      &   22 $\pm$ 1           &     19 $\pm$ 1
\\
log g (cgs)                               &   3.53 $\pm$ 0.01      &     3.52 $\pm$ 0.01 
\\
RV Semi-amplitude (km s$^{-1}$)           &   257 $\pm$ 7          &    343 $\pm$ 9
\\
mean Teff (K)                             &   35000 (Fixed)        &     29900 $\pm$ 700
 \\
Surface potential ($\Omega$)              &   3.14 $\pm$ 0.02      &     3.14 (Fixed)     
\\
M$_{Bol}$                                 &   -9.8 $\pm$ 1         &     -8.8 $\pm$ 1
\\
\hline

        \end{tabular}
   
\end{table}


\begin{figure}[h!]
   \centering  
 \label{fig:lcfit}
  \resizebox{\columnwidth}{!}{\includegraphics{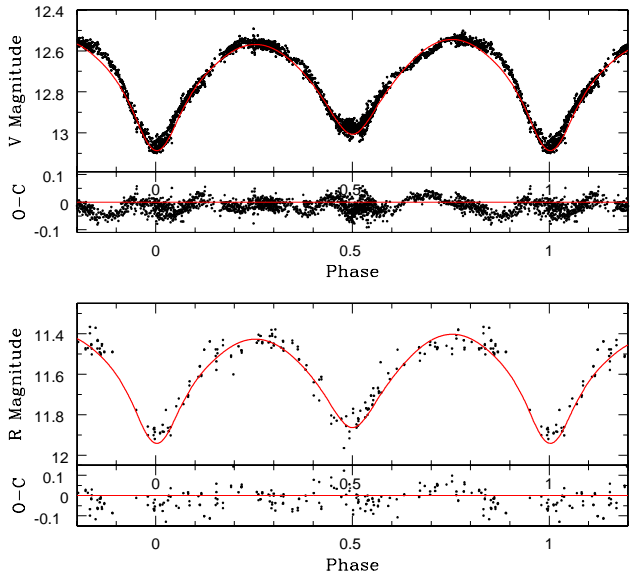}}
   \caption{The $V$ band and $R$ band light curves of B17 with the best fit model overlaid. 
The panels below the light curves show the deviations from the model.}
\end{figure}


\begin{figure}[!ht]
   \centering  
 \label{fig:rvfit}
  \resizebox{\columnwidth}{!}{\includegraphics[trim = 0mm 70mm 0mm 0mm, clip]{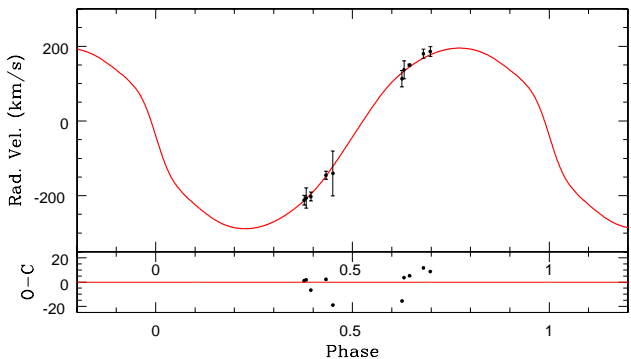}}
   \caption{The RV curve for the primary component obtained from the WHT data, with the best
 fit overlaid (solid line). The panel below shows the 
residuals between the observed data and the model.}
\end{figure}

\section {Discussion}

Given the difficulties in determining a unique model fit to the data (Sect. 5), we regard 
the parameters presented in Table~2 to be provisional. As a check on consistency we 
have calculated a  preliminary bolometric magnitude for the components, following a similar procedure
the that employed by \cite{neg08}. Assuming that both
 components are of similar colour\footnote{The colour differences between an O7 
and an O9 supergiant are $(J-V)_{0}=0.09$ and $(K-V)_{0}=0.11$ \citep{weg94}}, 
we adopted the 2MASS value for $(J-K_{{\rm S}})$ used in \cite{neg08},
 the effective temperature $T_{{\rm eff}}$ and bolometric correction (BC)
 calibrations of \cite{mar05}, and the intrinsic $(J-V)_{0}$ and
 $(K-V)_{0}$ colour calibrations of \cite{weg94}. Using the 2MASS observed
 $(J-K_{{\rm S}})$, we derived $E(J-K_S)$. The reddening of the system was
 calculated using the relation $A_{K_{{\rm S}}}=0.67E(J-K_{{\rm S}})$, due to the
 reddening in the association being close to standard \citep{han03}. The reddening
 in the $V$ band was calculated using the relation $0.112A_{V}$$\simeq$$A_{K}$
\citep{rie85}. The absolute $V$ band magnitude for the primary was then 
calculated, using the $V$ band value of the secondary minimum and adopting the 
distance modulus of 11.3, obtained by averaging spectroscopic distances
\citep{kim07}. A semi-observational $M_{{\rm bol}}$
was then calculated to be 
$-9.8\pm0.2$ and $-9.2\pm0.2$ for the primary and secondary, by adding $(V-K)_{0}$ and 
the BC to the $V_0$, assuming an uncertainty of one spectral type. With this value, the
 luminosities were calculated to be log($L_{1}/L_{\odot})=5.8\pm0.1$ and 
log($L_{2}/L_{\odot})=5.6\pm0.1$, noting that the main
 source of error is likely to be the uncertainty in the spectral type and hence 
temperature and BC. We regard these estimates as  upper limits since it assumes
 that both components are completely eclipsed during the minima.

 The positions of the two components of B17 in the HR diagram (Fig.~10) are consistent with other known members of 
 Cyg OB2 suggesting that it too is a {\em bona fide} member, with an age of $\sim$2.5Myr \citep{neg08}.
 
With a likely spectral type of O9~Iaf, the secondary appears slightly more
 evolved than the primary and hence was likely the initially more massive star,
 with both evolving from very early O main sequence stars. We note that a system
 composed of O7~Iaf \& O9~Iaf stars is also compatible with the other massive 
 evolved binaries in Cyg OB2. Indeed, the 6.6~day Ofpe/WNL+O6.5-7  binary Cyg OB2 $\#$5 
\citep{lin09} appears to be remarkably similar to B17, although with the Ofpe/WNL star 
being slightly more evolved than the O9~Iaf 
secondary in B17. The slightly longer period of  Cyg OB2 $\#$5 can be explained by one or more from
 (i) initial birth parameters, (ii) mass transfer from one star to the other and 
(iii) mass lost by both stars via stellar winds.

\cite{mar07,mar08} used observations of the Galactic Centre and Arches
 clusters to examine the evolutionary pathways of stars at galactic metallicity with masses in excess of
30M$_{\odot}$, suggesting the following evolutionary pathways \citep[also see][]{cro95}:\\
$\bullet$ { $\sim$30-60M$_{\odot}$ : O $\rightarrow$ Ofpe/WN9 $\rightleftharpoons$ LBV $\rightarrow$ WN8 $\rightarrow$ WN/C} 
\\
$\bullet$ { $\sim$60-120M$_{\odot}$ : O $\rightarrow$ Of $\rightarrow$ WNL + abs $\rightarrow$WN7} 
\\

With an age of $\sim$2.5 Myr and spectral classifications of O7 and O9, B17 appears to lie at the dividing point between the 2 evolutionary  pathways. 
 However, given that it 
appears significantly less massive than the WN6ha + WN6ha binary WR20a 
(m$_1$=m$_2$=83$M_{\odot}$ \citealt{rau04,bon04}) we suspect that
 it will instead evolve to resemble Cyg OB2 $\#$5 and hence to a configuration similar
to GCIRS 16SW, composed of two cooler extreme B supergiants/LBV candidates 
($P_{{\rm orb}}=19.45$ days; \citealt{mar06}),
 with stellar mass loss resulting in an eventual lengthening of the orbital period
 of B17. Indeed, assuming the system avoids merger during the LBV phase and, 
 remaining bound, receives a favourable SNe kick to reduce the orbital separation
 it might  briefly form a high mass X-ray binary with a WR mass donor prior to the second SN.

Irrespective of its ultimate fate, B17 is amongst the brightest/most massive
 systems in Cyg OB2 and adds to the increasing number of massive binaries 
identified within it (\citealt[][see Table 3]{kim09}). Similar trends for both a high binary fraction
 and the multiplicity of the brightest/most evolved cluster members  
have been observed for Pismis 24 \citep{mai08}, NGC3603 \citep{sch08},
Westerlund 1 (\citealt{cla08}; \citealt{rit09}) and potentially the Arches \citep{cla09}. 
If these trends continue
 it will have significant implications for the formation channels and 
relative production rates of both low and high mass X-ray binaries and systems
 comprising of two relativistic objects \citep{kob07}.


\begin{figure}[!ht]
   \centering   
 \label{fig:cygob2}
 \resizebox{\columnwidth}{!}{\includegraphics[trim = 0mm 0mm 0mm 0mm, clip]{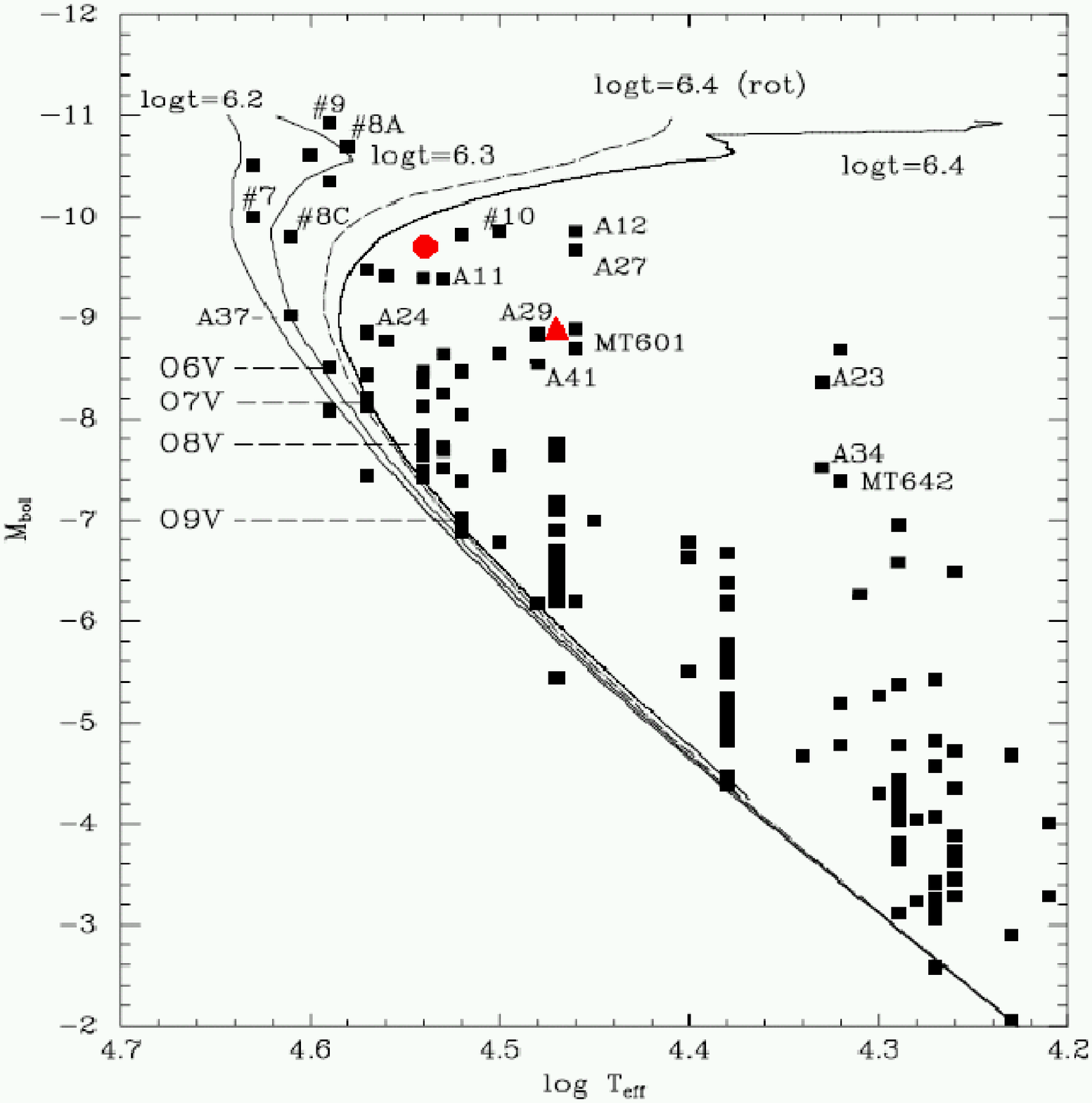}}
  \caption{Updated semi-observational HR diagram from \cite{neg08}, based on 
published spectral classes and 2MASS $JHK_{{\rm S}}$ photometry and a DM of 11.3. The continuous lines
 are non-rotating isochrones for $\log t=6.2$, $6.3$ and $6.4$ from \cite{sch92} and
 the dashed line is the $\log t=6.4$ isochrone in the high-rotation models from 
\cite{mey03}. The circle and the triangle show the positions of the primary and secondary components of B17.}
\end{figure}

\begin{table*}    
\caption {Evolved massive binaries in Cyg OB2 from Kiminki et al (2009).} 
\label{table:kiminki}
    \centering
        \begin{tabular}{l l l l}

\hline\hline	
Star                & 	Sp Types                   &  Period (days)      &    References \\
\hline 
MT05                &   O9\,III \& \it{mid B}     &  25.1399 (0.0008)   & \cite{kim09} \\
MT720               &   \it{early B \& early B}   &  $<$5               & \cite{kim09} \\
Schulte 3           &   O6\,IV \&O9\,III          &  4.7464 (0.0002)    & \cite{kim09}, Kinemuchi et al. in prep\\
Schulte 5 ($\#$5)   &   O7\,I \& Ofpe/WN9          &  6.6                & \cite{wil48}, \cite{wil51}, \cite{mic53} \\
                    &                             &                     & \cite{wal73}, \cite{con97}, \cite{rau99}   \\     
Schulte 8a ($\#$8a) &	O6\,If \& O5.5\,III(f)     &  21.9               & \cite{rom69}, \cite{deb04}                       \\
Schulte 9 ($\#$9)   &   O5.5\,If \&  O6-7           &  2.35 yrs           & \cite{naz08}                    \\
Schulte 73          &   O8\,II \& \it{O8?}        &  17.4 (0.2)         & \cite{kim09}  \\
B17                 & 	O7\,Ia \& O9\,I              &  4.0217 (0.0004)    &  This paper       \\
\hline			 
       
        \end{tabular}
    
\end{table*}

\section{Summary}

Using photometric and spectroscopic data, we have demonstrated that B17 is an
 eclipsing, double lined spectroscopic binary comprising two supergiants with 
preliminary classifications of O7Iaf and O9Iaf.   The spectra are highly variable, and with a subset revealing
 features from both stars, raise the possibility of achieving a dynamical mass
 determination for both components. Utilising both the photometric lighturve and our
limited RV dataset we attempted to determine an initial orbital solution for the binary.

Despite the 
morphology of the lightcurve indicating a semi-contact configuration we were unable to
to achieve convergence for such a hypothesis and hence were forced to adopted an over-contact configuration.
In the absence of a full RV curve for both system components we were forced to fix the binary mass ratio, and had to 
include the presence of a star spot to address the observed asymmetries in the lightcurve (which are likely due to the effects of 
binary mass transfer). However, we were
still unable to fully fit both secondary eclipse and the lightcurve between orbital phase $\sim$0.6-0.9
with such a model; as such we regard the modelling results presented in this work as provisional. We anticipate
 that a full RV curve for both components will be necessary to obtain more precise parameters  for the system; 
additional data to accomplish this goal are currently  being obtained and refined analysis will be 
presented in a future paper.

Nevertheless, the provisional distance calculation of 1.5-1.8 kpc obtained from the
 lightcurve analysis agrees with previously published values for the distance
 to Cyg OB2, being inconsistent with the distance of 900-950 pc 
determined by \cite{lin09} with the Cyg~OB2 \#5 light curve analysis.

When placed in the HR diagram, B17 appears to be consistent with  the age and 
stellar population of the Cyg OB2 association. Assuming the system avoids
 merger, it is likely to evolve through an extreme B 
supergiant/LBV phase into a long period WR+WR binary configuration 
as mass loss via stellar winds increases the orbital separation.
In combination with the recent work  of \cite{kim09} and \cite{kob07} the results of our analysis
 provides additional evidence that Cyg~OB2 has a very high fraction of massive binary stars. Such an observational constraint
 needs to be considered when determining the initial mass function of the association as it may  both influence the slope 
of the relationship and also  lead to a population of artificially
 massive stars, resulting in the inflation of a putative  high mass cut-off to the IMF.

\begin{acknowledgements}

We thank Pedro Pastor and Manuel M\'{e}ndez for having obtained and reduced the photometric data. Amparo Marco for help with the 2004 observing run and Miriam Garc\'{i}a for assistance with the 2006 INT run. We thank Dan Kiminki, Fraser Lewis and Chris Evans for useful discussions and reading of the manuscript. We also thank the referee for his guidance in completing the paper for publication. The Faulkes Telescope Project is an educational and research arm of the Las
 Cumbres Observatory Global Telescope Network (LCOGT). VS acknowledges support
 from the Dill Faulkes Educational Trust. 
This research is partially supported by the Spanish Ministerio de Ciencia e
 Innovaci\'on undergrants AYA2008-06166-C03-03 and Consolider-GTC CSD2006-70.
The G.D. Cassini telescope is operated at the Loiano Observatory by the
 Osservatorio Astronomico di Bologna. The WHT is operated on the island of
 La Palma by the Isaac Newton Group in the Spanish Observatorio del Roque 
de Los Muchachos of the Instituto de Astrof\'{\i}sica de Canarias. The 2006
 observations were taken as part of the service programme
(programme SW2005A20).

\end{acknowledgements}

\end{document}